# Nano-Raman spectroscopy of silicon surfaces

*P. G. Spizzirri[1], J.-H. Fang[1], S. Rubanov[2], E. Gauja[3] and S. Prawer[1].

[1]*ARC Centre of Excellence for Quantum Computer Technology*
*School of Physics, University of Melbourne, Melbourne, Victoria, 3010.*
[2]*Bio21 Molecular Science and Biotechnology Institute*
*University of Melbourne, Melbourne, Victoria, 3010.*
[3]*ARC Centre of Excellence for Quantum Computer Technology*
*School of Physics, University of New South Wales, Sydney, New South Wales, 2052.*

*pgspiz@unimelb.edu.au

Near-field enhanced, nano-Raman spectroscopy has been successfully used to probe the surface chemistry of silicon prepared using standard wafer cleaning and processing techniques. The results demonstrate the utility of this measurement for probing the local surface chemical nano-environment with very high sensitivity. Enhancements were observed for the vibrational (stretching) modes of Si-H, F-Si-H and possibly also B-O-Si consistent with the surface treatments applied. The nano-probes did not enhance the phononic features of the silicon substrate.

## 1 INTRODUCTION

Significant interest in the fabrication of nano-materials over the last 2 decades has seen the development of nano-balls [1], -tubes [2], -wires [3], -ribbons [4] and sheets [5] in a variety of materials including carbon and silicon. Interestingly, the motivation for fabricating structures of this size arises from their unique physical properties which allow them to be used in ways that their bulk counterparts cannot. Characterizing nano-systems, including surfaces however, still presents challenges with few techniques available that are sensitive enough to probe the few atom limit and still provide structural information.

Metallic nano-structures have found new aplications as candidates for nano-optics. With dimensions well below the wavelength of visible (excitation) light, they undergo oscillations in the surface electron charge density (i.e. surface plasmons) which are induced by the applied field. These surface modes represent the collective vibrations of an electron gas (i.e. plasma). When the dipolar plasmon resonance of the nano-particle is resonant with the excitation source, the particle will radiate (coherently) light characteristic of dipolar (and higher order) radiation. This results in a spatially defined, intensity enhancement from areas surrounding the nano-particle. The dipolar field decays with distance away from the particle so the effective field only extends tens of nanometers [6].

Surface plasmons on metal nano-particles can also be used to enhance the sensitivity of spectroscopic measurement techniques such as inelastic (Raman) scattering. Significant enhancements to the scattering cross section, sometimes by as much as $10^{15}$, make possible the detection of even single molecules [7]. The effect is known as surface enhanced Raman spectroscopy (SERS) and has been used widely for the study of adsorbates on metal nano-particles. The enhancement mechanism is still poorly understood but is thought to involve either: (i) classical electromagnetic field enhancment, (ii) the formation of a charge transfer complex or a combination of these effects.

Since the enhancement field is highly spatially localised, the effect has been successfully combined with scanning probe techniques such as atomic force (AFM) and scanning tunneling microscopies (STM). This combination offers high resolution vibrational spectroscopy coupled with spatial resolutions below the diffraction limit, even down to a few nanometers [7]. Termed tip enhanced Raman spectroscopy (TERS), the non-contact nature of the measurement and the topographical information it offers make this a very appealing tool. Unfortunately, the complexity and cost of the combined instrumentation (AFM/Raman) has limited the broad application of this technique to the spectroscopic study of surfaces.

In this work, we investigate the use of silver nano-particles as stationary, near-field, contact nano-probes for Raman measurements of thin surface layers and monolayers. Termed Probe Enhanced Raman Spectroscopy (PERS), this study differs from other investigations where (i) measurements are performed in solution and analytes are associated with the surface of the nano-probe to effect signal enhancement (SERS) or (ii) the nano-probe is attached to a scanning structure placed in proximity to a surface (TERS). Instead, we will be relying upon the dipolar field enhancement associated with each silver nano-particle to provide a locally enhanced area around each deposited nano-probe. While this approach does not offer any opportunity to spatially map the surface controllably, it does provide a very simple means of performing point measurements at various locations using only a micro-Raman spectrometer.

The two silicon surface preparations investigated in this work were produced using standard silicon processing techniques as employed in the fabrication of metal-oxide-semiconductor (MOS) devices. Aqueous treatments are still used in the first stage of controlling the surface chemistry and contamination levels of silicon wafers. This is an area of increasing importance given the current efforts to produce few electron, nano-devices. The insulating oxide grown on silicon for MOS architectures is also a crucial element for successful device operation. For this reason, the two surface terminations chosen for this nano-spectroscopic study are: (i) a hydrogen terminated silicon surface (monolayer) and (ii) a thin (< 10 nm) thermal oxide as used in MOS devices.

## 2 EXPERIMENTAL DETAILS

The nano-particle fabrication, surface preparation and measurements were all performed in a cleanroom environment to reduce contamination.

### 2.1 Electron Microscopy

Materials were characterised using an FEI Quanta 600 FEG scanning electron microscope (SEM) while high resolution imaging (electron transmission measurements) were performed using a FEI Tecnai TF20 transmission electron microscope (TEM) operating at 200 keV. Thin foils (preparation) for TEM measurement were prepared using a focussed ion beam.

### 2.2 Optical Spectroscopy

Room temperature Raman and PERS measurements were performed on a Renishaw RM1000 micro-Raman spectrometer with a 2.41 eV argon ion laser excitation source. The notch filter spectral response profile for this instrument prevented measurement below ~ 100 cm$^{-1}$. Non-enhanced measurements were performed with 10 mW of laser power and 180 s integration times in order to detect weaker signals. By contrast, the PERS measurements were generally performed with only 2 mW of laser power and 10-30 s integration times.

The optical properties and thermal oxide layer thickness were determined using a Jobin Yvon spectroscopic ellipsometer (SE). Measurements were taken from 250–620 nm in air at room temperature at an angle of incidence of 70 degrees. Silicon wafer quarters with thermally grown oxides were analyzed using the 3 mm x 1 mm optical beam. The parameters $\psi$ and $\Delta$, termed the ellipsometric angles, characterize the amplitude ratio and the phase difference between the two measured polarizations, respectively. The measurements reported in this work report values of Ic and Is which are related to the ellipsometric angles by:

$$Ic = \sin 2\Psi \cos \Delta \quad \text{and} \quad Is = \sin 2\Psi \sin \Delta$$

### 2.3 Synthesis of Silver Nanoparticles

Colloidal suspensions of silver nanoparticles were prepared by the method of Lee and Meisel using a silver nitrate precursor [8]. The preparation was drop cast onto the surface of silicon wafers which had been prepared with different surface chemistries.

### 2.4 Preparation of Silicon Surfaces

High resistivity silicon was cleaned using industry standard procedures to remove surface contaminants. The wet chemical processing included Pirana and RCA2 procedures which are well suited to hydrocarbon and metal ion (contamination) removal. The silicon native oxide layer was removed using a hydrofluoric acid (HF) etchant which also leaves the surface hydrogen terminated (i.e. a surface hydrogen monolayer) [9]. This surface preparation was investigated as prepared.

The second surface preparation studied was a high quality, thin oxide layer which was prepared using an ultra-dry thermal oxidation process. Pre-cleaned and hydrogen terminated silicon was heated in a quartz furnace at 800 °C for ~20 minutes in an ultra-dry $O_{2(g)}$ ambient. The thin oxide layers (<10 nm) produced were measured as prepared. Reagents used were generally of VLSI grade purity.

## 3 RESULTS AND DISCUSSION

An example of the silver nano-particles fabricated in this work can be seen in the SEM image of Figure 1. It is clear from this image that these particles tend to cluster together when produced and dispersed in this fashion on silicon surfaces. They also have a range of shapes and sizes with spheroids, rods and wires all evident in the figure. The spheroids have diameters ranging from 80 to 140 nm, the rods are ~40 to 50 nm wide by ~200 nm long and the wires are ~40 to 50 nm wide with lengths ranging from 400 nm to many microns. The PERS activity of these nano-particles when using 2.41 eV laser excitation was established through the observation of Raman vibrational modes from the fluorescent dye Rhodamine 6G adsorbed onto the nano-probe surfaces.

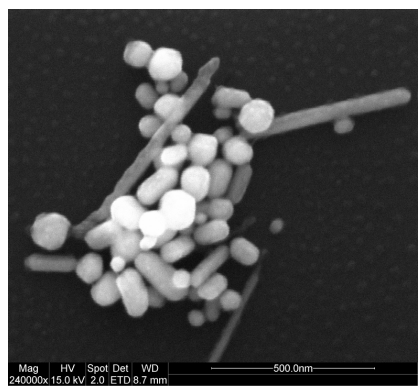

**Fig. 1.** A collection of the metallic silver, enhancement nano-particles dispersed on a silicon wafer.

Silicon, like diamond, has only one first order Raman active phonon located at the Brillouin-zone centre. The first order Raman (stokes) spectrum consists of one strong peak at 520 cm$^{-1}$ arising from the creation of the triply degenerate, long wavelength transverse optical phonon (TO). The second order spectrum is much weaker with features ranging from 0–1050 cm$^{-1}$. Higher order features, assigned as 3TO and 4TO scattering, are evident in the 1400–1600 cm$^{-1}$ and 1800–2000 cm$^{-1}$ region of the spectrum and combination bands can be found around 1000-1250 cm$^{-1}$. While this provides a rich spectral phononic landscape for silicon, the intensity of these bands make it more difficult to observe the weaker spectral features considered in this work.

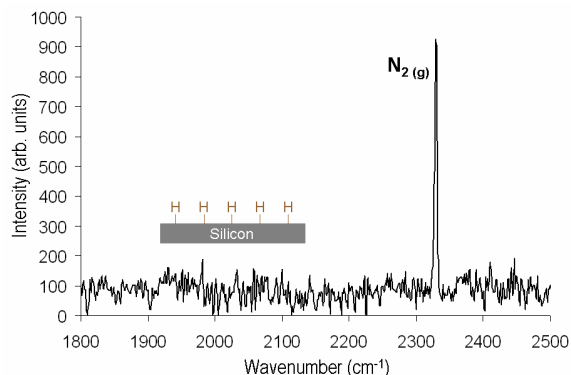

**Fig. 2.** Raman spectrum of a hydrogen terminated silicon surface without the enhancement nano-particles. The sharp transition at ~2330 cm$^{-1}$ arises from ambient nitrogen (molecular) gas (Q-branch).

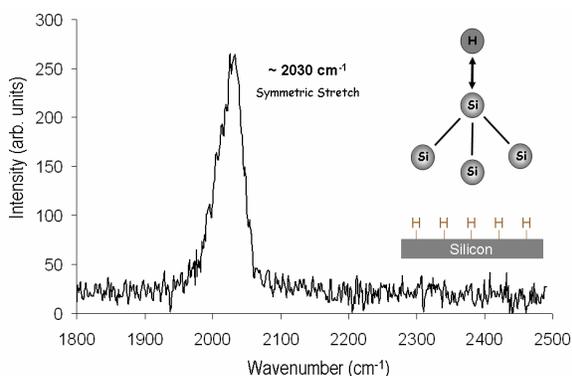

**Fig. 3.** Near-field enhanced Raman spectrum of a hydrogen terminated silicon surface. The transition at ~2030 cm$^{-1}$ is assigned as Si-H symmetric bond stretching.

For silicon prepared with a hydrogen terminated surface but without silver nano-particles (i.e. no enhancement), the spectrum below 1500 cm$^{-1}$ comprises mainly phononic silicon character. An example can be seen in the pink trace of Figure 4. Above 2000 cm$^{-1}$ however, there is a sharp transition arising from ambient nitrogen gas at 2332 cm$^{-1}$ (see Figure 2) [10]. Apart from this, there are no other features evident up to 3000 cm$^{-1}$.

After adding the silver nano-probes randomly to the Si-H terminated surface, a strong, clear signature peak for the silicon mono-hydride stretching (symmetric) mode is evident at ~2030 cm$^{-1}$ (see Figure 3) [11]. This transition appears to be assymetric and broad consisting of a number of overlapping bands. It has been suggested [11] that the linewidth of the Si-H stretching mode may be used to evaluate the silicon surface homogeneity. Since the mode frequency depends upon the specific bonding of the silicon to neighboring atoms, it can provide a sensitive measure of surface inhomogeneities. The linewidth observed in this work is much greater than the expected homogeneous linewidth (~1 cm$^{-1}$) suggesting that the defect density at this interface is relatively high. It should also be noted that the transition for molecular nitrogen is not observed in the enhanced spectrum because the measurement time and laser power are significantly reduced.

Other workers who have studied hydrogenated silicon using Raman scattering have also observed Si-H modes from a wide range [12-14] of modified silicon surfaces. These include: treated silicon using very high laser intensities [15] or highly roughened [16], amorphous [12], nanocrystalline [17] and ion implanted silicon where the density of Si-H states has been increased to levels far greater greater than that found in a surface monolayer. This work however, reports for the first time the near-field measurement of Si-H stretching modes from a planar, crystalline silicon interface.

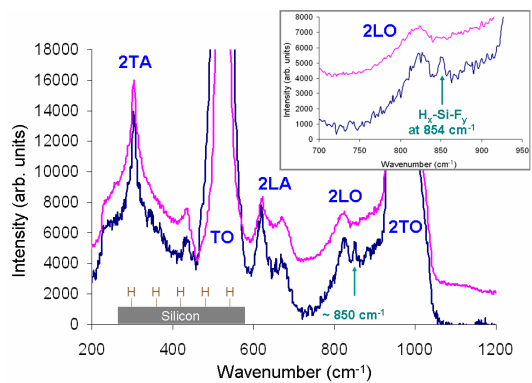

**Fig. 4.** Near-field enhanced Raman spectrum of hydrogen terminated silicon (blue trace) compared with the same surface treatment in the absence of enhancement (pink trace). Phononic silicon modes have been labeled as: transverse acoustic (TA), transverse optic (TO), longitudinal acoustic (LA) and longitudinal optic (LO) for clarity. The TO and 2TO bands are off scale on this plot and both measurements have been scaled to the intensity of the 2TA mode. The peak at ~ 850 cm$^{-1}$ is assigned as hydrogen associated silicon fluoride ($H_x$-Si-$F_y$). The inset spectrum, which is a magnified view, more clearly shows this band.

In addition to being able to detect Si-H modes from the hydrogenated surface, this measurement has also proven sensitive to the detection of hydrogen associated silicon fluoride ($H_x$-Si-$F_y$) which is thought to be a reaction intermediate that forms during the dissolution of silicon dioxide when using $HF_{(aq)}$. This species has a mode frequency around 850 cm$^{-1}$ [18] and was only observed at a few locations for the samples produced in this work (see Figure 4). Once again, this transition has not previously been observed from a monolayer at the silicon interface using Raman spectroscopy. It is somewhat surprising that this mode was observed since it is thought that HF etching does not usually lead to F-termination even though Si-F is thermodynamically more stable than Si-H [19]. Work by Fenner et al [20] has shown that surface residues of fluorine following similar treatments may be present at the level of approximately $5\times10^{-3}$ (± 0.002) of a monolayer.

The second controlled surface preparation investigated using near-field measurements was a high quality, thermal oxide layer on silicon. A high resolution TEM analysis of this material is shown in Fig. 5. This image depicts a high quality, sharp Si-SiO$_2$ interface with an oxide of approximately 7 nm thickness. The optical properties and thickness of this film were further investigated using spectroscopic ellipsometry with the data shown in Fig. 6. In order to analyze the ellipsometric measurements $Ic$ and $Is$, an appropriate physical model was needed to extract the optical properties of the SiO$_2$ layer. From the TEM data, a physical model of a 2 layer system was adopted with the first layer assigned as SiO$_2$ (~ 7 nm) and the second as bulk, crystalline silicon. Modeling the data over the entire spectral range measured using reference values for silicon and SiO$_2$ yielded an excellent fit (solid line) with a reported SiO$_2$ film thickness of 6.6 nm.

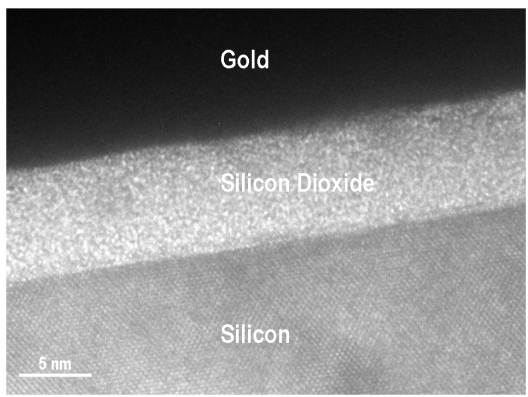

**Fig. 5.** Cross sectional TEM image of the thermally grown silicon oxide layer which is approximately 7 nm thick. The gold layer was deposited to protect the sample during FIB sample (foil) preparation.

The thermal SiO$_2$ layer was subsequently studied using micro-Raman spectroscopy to see if any Si-O vibrational modes could be observed. Many of the glass vibrations are known to occur between 850-1060 cm$^{-1}$ (Si-O stretching) and 400-700 cm$^{-1}$ (bridging Si-O-Si vibrations). These are regions where the silicon phononic contribution to the spectrum dominates [21, 22]. Indeed, in the absence of silver nano-particles, no significant spectral features attributable to the SiO$_2$ layer are observed.

Measurements were repeated with silver nano-probes located on the thermal oxide surface and a difference was noted with a new band emerging at only a few locations on the sample at around 1400 cm$^{-1}$. This data, along with a measurement from the same sample taken in the absence of the nano-probes, is shown in Figure 7. The narrow feature at around 1550 cm$^{-1}$ arises from ambient molecular oxygen and is stronger in the measurement without nano-probes because of the higher laser powers and longer measurement time used. The transition at ~1450 cm$^{-1}$ arises from third order scattering from silicon and is usually the dominant feature in this region with only minor contributions arising from silicon combination bands between 1000 and 1250 cm$^{-1}$.

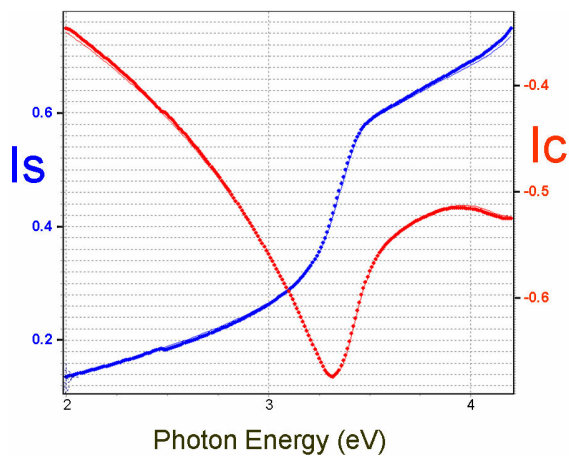

**Fig. 6.** Spectroscopic ellipsometry meaurements (dots) of the Si-SiO$_2$ sample modelled (solid lines) as a two layer system. The quality of the model fit is evidenced by its overlap with the measured data points over the entire spectral range.

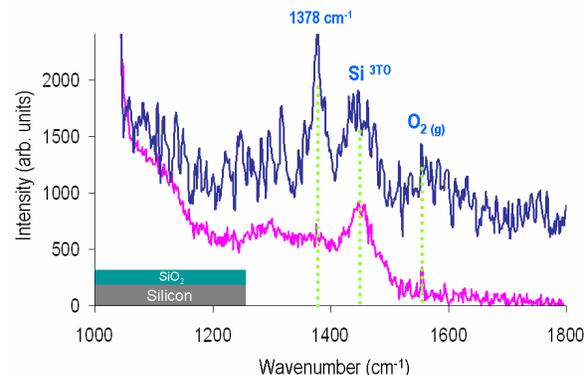

**Fig. 7.** Near-field enhanced Raman spectrum of SiO$_2$ (thermal) terminated silicon. The band at ~1450 cm$^{-1}$ is assigned as the third order silicon TO phonon, the ~1550 cm$^{-1}$ band arises from molecular (ambient oxygen) and the ~1380 cm$^{-1}$ band may originate from the stretching vibration of an Si-O-B bridging structure.

The new transition emerging at 1378 cm$^{-1}$ in the enhanced spectrum is more intense than the third order silicon band. This transition, which has not been observed from thin SiO$_2$ layers on silicon before, may arise from boron incorporation into the oxide. In low-boron-content borosilicate glasses [23] and condensed Vycor [24], a transition attributed to stretching modes of a B–O–Si bridging structure has been observed at 1380 cm$^{-1}$. In this work, the oxide has been grown through the sacrificial oxidation of the supporting silicon wafer. So while it is conceivable that this transition has arisen from B-O-Si stretching modes, it is surprising that it can be measured since the source material was only very lightly doped with boron ($<10^{14}$ cm$^{-3}$). It is also known however, that boron segregates into surface oxides during thermal treatment which would result in a significant increase in the SiO$_2$ boron content [25]. Nonetheless, this may be a significant result if the origin of this transition is a boron species in the oxide since these materials are fundamental to the production of metal-oxide-semiconductors and new tools to aid in their characterization would be welcomed.

## 5 CONCLUSIONS

These measurements, which represent sampling from random locations on the silicon and silicon dioxide surface, have demonstrated that near-field signal enhancements from silver nano-particles are demonstrable. Given the limited range of the enhancement field (i.e. tens of nanometers) and the size of the nanoparticles used in this work, the signals which have been observed from monolayers and thin films must have undergone significant enhancement. These results compare well with other work where larger samples have been measured.

This work has also shown that it is the enhancement of vibrational modes of interface states rather than bulk, phononic enhancements from the substrate or thin oxide that dominate in this type of near-field measurement.

## 8. ACKNOWLEDGMENTS

This work was supported by the Australian Research Council.